\documentclass[11pt]{article}

\usepackage[usenames,dvipsnames]{xcolor}
\definecolor{Gred}{RGB}{219, 50, 54}
\definecolor{ToCgreen}{RGB}{0, 128, 0}

\usepackage[margin=1in]{geometry}

\usepackage[latin1]{inputenc}
\usepackage[T1]{fontenc}
\usepackage[scale=0.97]{XCharter} 

\usepackage[libertine,bigdelims,vvarbb,scaled=1.05]{newtxmath} 

\usepackage{babel}
\usepackage[spacing=true,kerning=true,babel=true,tracking=true]{microtype}

\DeclareMathAlphabet{\pazocal}{OMS}{zplm}{m}{n} 
\renewcommand{\mathcal}[1]{\pazocal{#1}}

\usepackage{makecell}
\usepackage{dsfont}

\usepackage{multirow}
\usepackage{amsmath,amsthm}
\usepackage{bm}
\usepackage{bbm}
\usepackage{textgreek}
\usepackage{mathtools}
\usepackage{enumitem}
\usepackage[numbers,comma,sort&compress]{natbib}
\usepackage{authblk}
\usepackage{graphicx}
\usepackage[font=small]{caption}
\usepackage[labelformat=simple]{subcaption}

\usepackage{float}
\usepackage[linesnumbered,ruled,vlined]{algorithm2e}
\SetKwInput{KwInput}{Input}
\SetKwInput{KwOutput}{Output}
\SetKwInOut{Promise}{Promise}
\SetKwInput{Goal}{Goal}
\SetKwProg{Fn}{function}{}{}
\SetKwFor{RepTimes}{repeat}{times}{}
\SetKwFunction{Ver}{Verify}
\SetKwFunction{Prep}{PrepareState}
\usepackage{algorithmic}

\usepackage{physics}
\usepackage{footnote}
\usepackage{xcolor}
\usepackage{mathrsfs}
\usepackage{bbm}
\usepackage{braket}
\usepackage[colorlinks]{hyperref}
\usepackage{cleveref}
\hypersetup{
      colorlinks=true,
  citecolor=ToCgreen,
  linkcolor=Sepia,
  filecolor=Gred,
  urlcolor=Gred
  }
\usepackage{placeins}

\newtheorem{theorem}{Theorem}[section]

\newtheorem{problem}[theorem]{Problem}

\newcommand{\algo}[1]{\hyperref[algo:#1]{Algorithm~\ref*{algo:#1}}}

\renewcommand{\leq}{\leqslant}

\renewcommand{\le}{\leqslant}

\newcommand{\cP}{\mathcal{P}}

\usepackage{textgreek}

\renewcommand{\tr}{\mathrm{tr}}
\renewcommand{\Tr}{\mathrm{tr}}

\newcommand{\0}{\mathbf{0}}

\def\Tr{\operatorname{tr}}\def\:{\hbox{\bf:}}

\renewcommand{\epsilon}{\varepsilon}

\allowdisplaybreaks

\begin{document}

\title{Adaptivity can help exponentially for shadow tomography}

\author{
Sitan Chen
\thanks{SEAS, Harvard University. Email: \href{mailto:sitan@seas.harvard.edu}{sitan@seas.harvard.edu}.}
\qquad\qquad
Weiyuan Gong
\thanks{SEAS, Harvard University. Email: \href{mailto:wgong@g.harvard.edu}{wgong@g.harvard.edu}.}
\qquad\qquad
Zhihan Zhang
\thanks{IIIS, Tsinghua University. Email: \href{mailto:zhihan-z21@mails.tsinghua.edu.cn}{zhihan-z21@mails.tsinghua.edu.cn}.}
}

\date{}
\maketitle

\begin{abstract}
In recent years there has been significant interest in understanding the statistical complexity of learning from quantum data under the constraint that one can only make unentangled measurements. While a key challenge in establishing tight lower bounds in this setting is to deal with the fact that the measurements can be chosen in an adaptive fashion, a recurring theme has been that adaptivity offers little advantage over more straightforward, nonadaptive protocols.
    
In this note, we offer a counterpoint to this. We show that for the basic task of shadow tomography, protocols that use adaptively chosen two-copy measurements can be exponentially more sample-efficient than any protocol that uses nonadaptive two-copy measurements.
\end{abstract}


\section{Introduction}
The ability to extract information from quantum data lies at the heart of quantum information science.
While entangled measurements are powerful for this purpose, practical constraints often necessitate the use of unentangled or restricted entangled measurements.
This restriction introduces a dichotomy in algorithms: adaptive algorithms can dynamically decide how to measure a new batch of input based on the previous history, while nonadaptive algorithms fix their measurement strategies ahead of time.
Establishing tight sample complexity lower bounds is challenging in the adaptive setting.

Surprisingly, for a wide range of quantum learning tasks, adaptivity provably offers no improvement and one can achieve optimal statistical rates using nonadaptive protocols.
Examples include quantum state tomography with respect to trace distance~\cite{haah2016sample,odonnell2016efficient,chen2022tight,kueng2017low}, shadow tomography with unentangled measurements~\cite{huang2020predicting,chen2022exponential}, identity testing and mixedness testing~\cite{bubeck2020entanglement,chen2022toward,chen2022exponential}, and purity testing~\cite{chen2022exponential,gong2024sample,anshu2022distributed}. Prior to this note, the only known example where adaptivity helps for quantum learning was for quantum state tomography with respect to infidelity $\gamma$ using unentangled measurement, where adaptive protocols require $O(2^{3n}/\gamma)$ copies~\cite{chen2022tight} while nonadaptive protocols require $\Theta(2^{3n}/\gamma^2)$ copies~\cite{haah2016sample}. Here we ask:
\begin{center}
{\em Are there natural quantum learning tasks where adaptive measurement strategies significantly outperform nonadaptive ones?}
\end{center}

We answer this in the affirmative by showing that adaptivity can help exponentially for the task of Pauli shadow tomography using two-copy measurements.
The problem can be formalized as follows:

\begin{problem}[Pauli shadow tomography]\label{prob:pauli_shadow}
Given access to copies of an $n$-qubit unknown state $\sigma$, as well as a subset $A$ of the Pauli group $\cP_n$, the goal is to estimate the expectation values $\{\tr(P\sigma)\}_{P\in A}$ to within $\epsilon$ additive error.
\end{problem}

\noindent This problem arises naturally in various contexts like the variational quantum eigensolver, quantum resource theory, quantum process tomography, and stabilizer state learning. A review of this literature is beyond the scope of this note and the reader can
refer to the related works section of~\cite{chen2024optimal} for a more detailed overview. 

Pauli shadow tomography provides a clean testbed for exploring statistical overheads incurred by near-term device constraints when learning quantum states~\cite{chen2022exponential,huang2022quantum,huang2021information} and channels~\cite{chen2024tight,chen2022quantum,chen2023efficientPauli,caro2024learning}. For instance, it is known that $\Theta(2^n/\epsilon^2)$ copies are necessary and sufficient for protocols that can only perform unentangled measurements~\cite{chen2022exponential}, even adaptively chosen ones. On the other hand, there is a protocol which performs Bell basis (nonadaptive) measurements on two copies of $\sigma$ at a time which can estimate $\abs{\tr(P\sigma)}$ for all $P\in\cP_n$ using only $O(n/\epsilon^4)$ copies. Recently, \cite{chen2024optimal} and~\cite{king2024triply} improved upon this to give a protocol with the same sample complexity which performs two-copy measurements and can estimate $\tr(P\sigma)$ rather than just their absolute values. 

Importantly, the latter protocol of~\cite{chen2024optimal,king2024triply} is adaptive: in the first step, it uses the protocol of~\cite{huang2021information} to estimate the absolute values, and in the second, it uses these absolute values to prepare a certain single-copy measurement to learn the signs. Given the absence of strong separations between what can be done adaptively versus nonadaptively, it is reasonable to conjecture that there is a nonadaptive alternative to the protocol in~\cite{chen2024optimal,king2024triply}.

\begin{theorem}\label{thm:main}
Any protocol that estimates $\tr(P\sigma)$ for all $P\in\cP_n$ to additive error less than $1$ using nonadaptive two-copy measurements requires $\Omega(2^{n/2})$ copies of $\sigma$. On the other hand, there is a protocol making adaptively chosen two-copy measurements that only uses $O(n)$ copies.
\end{theorem}

\noindent To our knowledge, this is the first result that proves an \emph{exponential} separation between adaptive and nonadaptive protocols in quantum learning, answering a question posed by Aharonov, Cotler, and Qi~\cite{aharonov2022quantum}.
The second half of Theorem~\ref{thm:main} follows from~\cite{chen2024optimal,king2024triply}. We provide the proof for the first half, i.e. the lower bound for nonadaptive protocols, in \Cref{sec:proof} below.

\section{Proof of the main result}
\subsection{Preliminaries}
We begin by recalling some standard notations in quantum information. An $n$-qubit quantum state can be written as a positive semi-definite matrix $\sigma\in\mathbb{C}^{2^n\times 2^n}$ with $\tr(\sigma)=1$. When the state has rank $1$ and thus $\tr(\sigma^2)=1$, it is a pure state and can be denoted as $\ket{\psi}$.  

The \emph{$n$-qubit Pauli group} $\cP_n\coloneqq \{I, X, Y, Z\}^{\otimes n}$ is the set of $n$-qubit \emph{Pauli strings}, where 
\begin{equation*}
    I=\begin{pmatrix}1 & 0\\0& 1\end{pmatrix}\,, \qquad X=\begin{pmatrix}0 & 1\\1& 0\end{pmatrix}\,, \qquad Y=\begin{pmatrix}0 & -i\\i& 0\end{pmatrix}\,, \qquad Z=\begin{pmatrix}1 & 0\\0& -1\end{pmatrix}
\end{equation*} 
are single-qubit Pauli matrices. We will need the following fact for the sum of products of pairs of Pauli matrices (see e.g. ~\cite[Lemma 4.10]{chen2022exponential} for proof):
\begin{align}\label{eq:pauli_sum}
\sum_{P\in\cP_ n}P\otimes P=2^n\,\mathrm{SWAP}_n\, ,
\end{align} 
where $\mathrm{SWAP}_n$ is the $2n$-qubit SWAP operator.

A quantum measurement $\mathcal{M}$ can be represented as a set of positive operator-valued measures (POVMs), i.e. a set of positive-semidefinite matrices $\{F_s\}_s$ satisfying $\sum_s F_s=I$. Here, each $F_s$ is a \emph{POVM element} corresponding to a \emph{measurement outcome} $s$. When we measure a quantum state $\sigma$ using the POVM $\{F_s\}_s$, the probability of observing outcome $s$ is $\Tr(F_s\sigma)$. When $F_s$ is rank-$1$ for all $s$, it is a \emph{rank-1 POVM}.~\cite[Lemma 4.8]{chen2022exponential} shows that any POVM can be information-theoretically simulated by a rank-$1$ POVM and classical post-processing. In the following, we will thus assume without loss of generality when we prove our lower bounds that all measurements in the learning protocols we consider only use rank-$1$ POVMs. For measurements over two copies, these can be written as
\begin{align*}
\{w_s\ket{\psi_s}\bra{\psi_s}\}_s\,,
\end{align*}
for $2n$-qubit pure states $\{\ket{\psi_s}\}$ and nonnegative weights $w_s$ with $\sum_s w_s = 2^{2n}$.

\subsection{Proof of the nonadaptive lower bound}\label{sec:proof}

To prove the nonadaptive lower bound in \Cref{thm:main}, we consider a distinguishing task in which we are given access to copies of an unknown $n$-qubit quantum state $\sigma$ and want to distinguish between two cases:
\begin{itemize}
    \item $\sigma$ is sampled from $\Bigl\{\sigma_a^+=\frac{I+P_a}{2^n}\Bigr\}_{a=1}^{4^n-1}$ uniformly for $P_a\in\cP_n\backslash I$. 
    \item $\sigma$ is sampled from $\Bigl\{\sigma_a^-=\frac{I-P_a}{2^n}\Bigr\}_{a=1}^{4^n-1}$ uniformly for $P_a\in\cP_n\backslash I$.
\end{itemize}
\noindent If we have a protocol that can solve \Cref{prob:pauli_shadow} for all $P\in\cP_n$ to additive error less than $1$ with high probability, then we can solve this distinguishing task using the same protocol. It thus suffices to prove a lower bound on the number of copies of $\sigma$ needed to solve the distinguishing problem with high probability.

For any nonadaptive protocols using at most $T$ two-copy measurements, we can assume that the protocol nonadaptively picks the POVM measurement $\mathcal{M}_t=\{w_s^t\ket{\psi_s^t}\bra{\psi_s^t}\}_s$, where each $\ket{\psi_s^t}$ is a $2n$-qubit pure state and $\sum_{s}w_s^t=2^{2n}$, in the $t$-th iteration. Given an $a\in\{1,...,4^n-1\}$, we denote by $p^+_{a,t}$ and $p^-_{a,t}$ the probability distributions over outcomes $s_t$ from measuring $\sigma_a^+$ and $\sigma_a^-$ respectively using $\mathcal{M}_t$. We also denote by $p^+_a$ and $p^-_a$ the probability distribution over the $T$ measurement outcomes $s_1,...,s_T$ for measuring $\sigma_a^+$ and $\sigma_a^-$.

By Le Cam's lemma~\cite{yu1997assouad} for hypothesis testing, if the total variation distance between the average probability distributions $\mathbb{E}_a[p^+_a]$ and $\mathbb{E}_a[p^-_a]$ is bounded by $o(1)$ for any nonadaptive sequence of two-copy POVM measurements $\{\mathcal{M}_t\}_{t=1}^T$, i.e.,
\begin{align*}
\text{TV}(\mathbb{E}_a[p^+_a],\mathbb{E}_a[p^-_a])\leq o(1),
\end{align*}
then any nonadaptive protocols using at most $T$ two-copy measurements cannot distinguish between these two cases with high probability. 

Note that for a sequence of nonadaptive measurements, the probability distribution $p^+_a$ and $p^-_a$ can be written as a tensor product of the probability distributions for each individual measurement $p^+_{a,t}$ and $p^-_{a,t}$. We thus have
\begin{align*}
\text{TV}(\mathbb{E}_a[p^+_a],\mathbb{E}_a[p^-_a])&\leq \mathbb{E}_a[\text{TV}(p^+_a,p^-_a)]=\mathbb{E}_a\Bigl[\text{TV}\Bigl(\bigotimes_{t=1}^Tp^+_{a,t},\bigotimes_{t=1}^Tp^-_{a,t}\Bigr)\Bigr]\leq \mathbb{E}_a\Bigl[\sum_{t=1}^T\text{TV}(p^+_{a,t},p^-_{a,t})\Bigr]\\
&=\sum_{t=1}^T\mathbb{E}_a[\text{TV}(p^+_{a,t},p^-_{a,t})]\leq T\max_{\text{two copy }\mathcal{M}_t}\mathbb{E}_a[\text{TV}(p^+_{a,t},p^-_{a,t})]
\end{align*}
where the first line follows from Jensen's inequality, the nonadaptivity of the protocol, and triangle inequality. 

It thus suffices to bound the average total variation distance for a \emph{single} two-copy measurements. For any rank-1 POVM $\mathcal{M}_t$ over two copies, we have
\begin{align*}
\MoveEqLeft\mathbb{E}_a[\text{TV}(p^+_{a,t},p^-_{a,t})]\\
&=\mathbb{E}_a\Bigl[\frac{1}{2}\sum_s\Bigl|\tr\Bigl[\Bigl(\frac{I+P_a}{2^n}\Bigr)^{\otimes2}w_s^t\ket{\psi_s^t}\bra{\psi_s^t}\Bigr]-\tr\Bigl[\Bigl(\frac{I-P_a}{2^n}\Bigr)^{\otimes2}w_s^t\ket{\psi_s^t}\bra{\psi_s^t}\Bigr]\Bigr|\Bigr]\\
&=\mathbb{E}_a\Bigl[\frac{1}{2^{2n}}\sum_sw_s^t\,\abs{\tr[(I\otimes P_a+P_a\otimes I)\ket{\psi_s^t}\bra{\psi_s^t}]}\Bigr]\\
&\leq\frac{1}{2^{2n}}\mathbb{E}_a\Bigl[\sqrt{\sum_{s}w_s^t\,\tr^2[(I\otimes P_a+P_a\otimes I)\ket{\psi_s^t}\bra{\psi_s^t}]\cdot\sum_{s}w_s^t}\Bigr]\\
&\le\frac{1}{2^n}\sqrt{\mathbb{E}_a\Bigl[\sum_{s}w_s^t\,\tr^2\left[(I\otimes P_a+P_a\otimes I)\ket{\psi_s^t}\bra{\psi_s^t}\right]\Bigr]}\\
&=\frac{1}{2^n}\sqrt{\mathbb{E}_a\Bigl[\sum_sw_s^t\bra{\psi_s^t}\bra{\psi_s^t}(I\otimes P_a+P_a\otimes I)^{\otimes 2}\ket{\psi_s^t}\ket{\psi_s^t}\Bigr]}\\
&=\frac{1}{2^n}\sqrt{\frac{1}{2^{2n}-1}\sum_sw_s^t\bra{\psi_s^t}\bra{\psi_s^t}\left(2^n(\text{SWAP}_{1,3}+\text{SWAP}_{1,4}+\text{SWAP}_{2,3}+\text{SWAP}_{2,4})-4I^{\otimes 4}\right)\ket{\psi_s^t}\ket{\psi_s^t}}\\
&\leq\frac{1}{2^n}\sqrt{\frac{1}{2^{2n}-1}\sum_sw_s^t(4\cdot 2^n-4)}\\
&=\frac{1}{2^n}\sqrt{\frac{4\cdot 2^{2n}}{2^{n}+1}}= O\Bigl(\frac{1}{2^{n/2}}\Bigr)
\end{align*}
where the third step follows from the Cauchy-Swartz inequality, the fourth step follows from the fact that $\sum_sw_s^t=2^{2n}$ and Jensen's inequality, and the sixth step follows from Eq.~\eqref{eq:pauli_sum} (here $\text{SWAP}_{i,j}$ denotes the $\text{SWAP}$ operator on the $i$-copy and the $j$-th copy). Therefore, we have
\begin{align*}
\text{TV}(\mathbb{E}_a[p^+_a],\mathbb{E}_a[p^-_a])\leq O(T\cdot 2^{-n/2}).
\end{align*}
This indicates that any nonadaptive protocols with $T\leq o(2^{n/2})$ two-copy measurement can not solve this distinguishing task with high probability, which yields the $\Omega(2^{n/2})$ lower bound in \Cref{thm:main} for two-copy nonadaptive protocols for solving \Cref{prob:pauli_shadow} for $\cP_n$.

\section{Discussion}

In this note, we proved the first exponential separation in sample complexity between learning protocols with and without adaptivity in the context of shadow tomography. Looking ahead, it would be interesting to study more fine-grained tradeoffs between sample complexity and adaptivity, e.g. as measured by the number of \emph{rounds} of adaptivity.
In the setting studied in this paper, one round of adaptivity is sufficient to obtain optimal dependence on $n$~\cite{chen2024optimal,king2024triply}. Are there natural quantum learning tasks where protocols with more rounds of adaptivity are strictly more powerful?

Another important question is to understand the computational complexity of optimally choosing adaptive measurement bases in quantum learning. While the adaptive protocols in~\cite{chen2024optimal,king2024triply} are sample-efficient, their adaptive choice of measurement currently requires at least exponential classical computation. Is this inherent, or can one prove a separation between nonadaptive and adaptive \emph{computationally efficient} protocols?


\section*{Acknowledgments}
We thank Hsin-Yuan Huang for helpful discussions and for encouraging us to write up this finding. We thank Jordan Cotler and Jerry Li for providing feedback on this manuscript, and Robbie King and Qi Ye for helpful discussions.


\bibliographystyle{MyRefFont}
\bibliography{NoteRef}

\end{document}